\documentclass[11pt,twoside]{article}

\usepackage{asp2004}
\usepackage{epsf}
\usepackage{lscape}

\begin{document}
\title{Evaporation and Thermal Balance of Tiny HI Clouds}
\author{Jonathan D.\ Slavin}
\affil{Harvard-Smithsonian Center for Astrophysics}

\begin{abstract}
We discuss the thermal evaporation of tiny H\,\textsc{i} clouds in the
interstellar medium. Cold neutral clouds will take $\sim 10^6 - 10^7$ yr to
evaporate if they are embedded in warm neutral gas and about half as long if
embedded in hot gas.  Heat flux saturation effects severely reduce the
evaporation rate of cold neutral medium (CNM) clouds into hot gas.  For
CNM clouds embedded in warm neutral medium (WNM) the much lower conductivity
results in slower evaporation.  This mass loss rate could still be
significant, however, if the environment is relatively quiescent.  Partial
ionization of the WNM gas would substantially reduce the conductivity and
lengthen the lifetime of the tiny H\,\textsc{i} clouds.  The ultimate
importance of thermal conduction to cloud evolution will depend on the role of
turbulence and the characteristics of the medium in which the clouds are
embedded.
\end{abstract}

\section{The Basics of Cloud Evaporation}

Cloud evaporation in the interstellar medium (ISM) is theorized to occur in
regions where a cool cloud is embedded in a warmer medium.  Thermal conduction
causes heat to flow into the cloud raising its temperature and pressure at the
boundary resulting in an evaporative outflow.  The energy equation that
governs this process is
\begin{equation}
\frac{\partial}{\partial t}\left(\frac{3}{2} P + \frac{1}{2} \rho v^2\right)
+ \nabla \cdot \left[\left(\frac{5}{2} P + \frac{1}{2} \rho v^2\right)\vec{v} 
+ \vec{q} \right] = -\Lambda,
\label{eq:energy}
\end{equation}
where $\Lambda$ is the net cooling rate, $\vec{q}$ is the heat flux, $P$ is
the thermal pressure, $\vec{v}$ is the velocity and $\rho$ is the density.
If externally generated changes to the system are slow compared with the
outflow speed, we may assume a steady flow.  If we also assume spherical 
symmetry and subsonic outflow ($v \ll c$, generally a good assumption) then
equation (\ref{eq:energy}) can be written as
\begin{equation}
\frac{d}{dr}\left(\frac{5}{2}\frac{P}{\rho} \dot{m} + 4\pi r^2 q\right) =
-\Lambda 4\pi r^2,
\end{equation}
where $\dot{m} = 4\pi \rho v r^2$ is the (constant) mass loss rate from the
cloud.  As can be seen from this equation, cooling effectively reduces heat
flux into the cloud while heating increases it.

\section{Thermal Conductivity in Neutral and Ionized Gas}

\begin{figure}[!ht]
\plotone{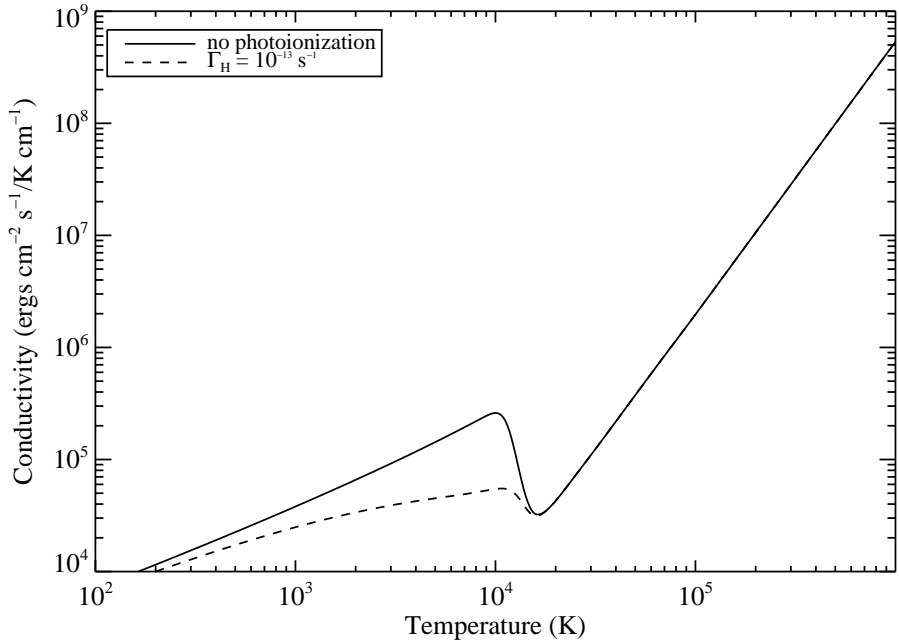}
\caption{Conductivity as a function of temperature for collisionally and
photoionized gas.  The solid line shows the conductivity for the case in which
only collisions ionize the gas and the dashed line shows the effects of having
a photoionization rate of 10$^{-13}$ s$^{-1}$ in gas at a pressure of 3000
cm$^{-3}$ K. The conductivity in gas in which the heat carrier is primarily
neutral H is strongly affected by the degree of ionization in the gas because
the presence of H$^{+}$ in the gas severely reduces the mean free path of the
neutral H atoms.\label{fig:ioneff}}
\end{figure}

The heat flux can generally be expressed as $\vec{q} = -\kappa \nabla T$
where $\kappa$ is the conductivity.  The conductivity depends strongly on both
the temperature and ionization in the gas.  Generally $\kappa \sim
n\bar{v}\ell \sim T^{1/2}/\sigma$ where $n$ is the gas density, $\bar{v}$ is
the mean speed of gas particles, $\ell$ is the mean free path and $\sigma$ is
the collision cross section.  The mean free path and collision cross section
here apply to the carrier of the heat in the plasma -- electrons in ionized
gas and H atoms in neutral gas.  For electrons in ionized gas $\sigma \sim
T^{-2}$ which leads to $\kappa \sim T^{5/2}$, the classical Spitzer
conductivity.  In neutral gas, heat is conducted primarily by elastic
collisions between H atoms with a cross section that goes roughly as
$T^{-0.3}$ \citep[see][]{Nowak+Ulmschneider_1977} leading to $\kappa \sim
T^{0.8}$.  However, when H$^+$ is present, the large H$^+$-H charge transfer
cross section dramatically reduces the mean free path and thus the
conductivity.  Figure \ref{fig:ioneff} demonstrates this by showing the
conductivity in gas that is subject only to collisional ionization (solid
line) compared to gas in which about a 20\% H ionization fraction is
maintained at low temperatures via an ionization rate of $\Gamma_\mathrm{H} =
10^{-13}$ s$^{-1}$ (at a pressure of $P/k_\mathrm{B} = 3000$ cm$^{-3}$ K)
(dashed line).  Clearly the ionization makes a dramatic difference at
temperatures of $T \sim 10^3- 10^4$ K such as are found in WNM/WIM gas.

\section{Heat Flux Saturation}

Another important factor in limiting the heat flux under certain circumstances
is what is known as saturation of the flux.  This comes about because no
matter how large a temperature gradient exists in the gas, the heat can only
flow at the rate at which it can be carried by either the electrons (in
ionized gas) or the H atoms.  This limitation can be expressed as
\begin{equation}
|q| \le q_\mathrm{sat} \approx \frac{3}{2} \rho c^3,
\end{equation}
where $c$ is the sound speed or thermal speed of the heat carrying particles.
For evaporation into hot ($T\sim10^6$ K) gas, saturation is important if the
saturation parameter, $\sigma_0$ \citep[see][]{Cowie+McKee_1977}, is greater
than 1. This parameter can be expressed as 
\begin{equation}
\sigma_0 = 3.226\frac{(T_f/10^6\,\mathrm{K})^3}{(P/10^4\,k_\mathrm{B})
R_\mathrm{cl}(\mathrm{pc})},
\end{equation}
where $R_\mathrm{cl}$ is the cloud radius and $T_f$ is the hot gas temperature
(the asymptotic temperature in the \citet{Cowie+McKee_1977} model of steady
cloud evaporation).  Saturation, which is most important for for high $T_f$,
low $P$ and for small clouds, can lead to an evaporation rate that is
substantially below the ``classical'' rate,
\begin{equation}
\dot{m}_{\mathrm{cl}}(\mathrm{M}_{\sun}/\mathrm{Myr}) = 0.436 (T_f/10^6
\mathrm{K})^{5/2} R_\mathrm{cl}(\mathrm{pc}).
\end{equation}
For $\sigma_0 \ga 5$, which should apply to the tiny H\,\textsc{i} clouds, the
factor by which the mass loss rate is reduced becomes $w \approx 1.22
\sigma_0^{-5/6}$.

\section{Thermal Balance and Radiative Cooling}

Radiative cooling can also act to reduce the evaporation rate for clouds.  The
strongest effects of cooling occur in the opposite regime from saturation
effects -- i.e.\ for high $P$, low $T_f$ and for large $R_\mathrm{cl}$.  
When cooling is large enough, evaporation is stopped altogether and the cloud
is radiatively stabilized.  With even more cooling, the cloud will condense.
For a given $P$ and $T$, one can find the critical cloud radius for
stabilization -- larger clouds condense, smaller ones evaporate.
Plane parallel clouds under these assumptions should \emph{always} condense
and elongated clouds should tend to become spherical.

\begin{figure}[!ht]
\plotone{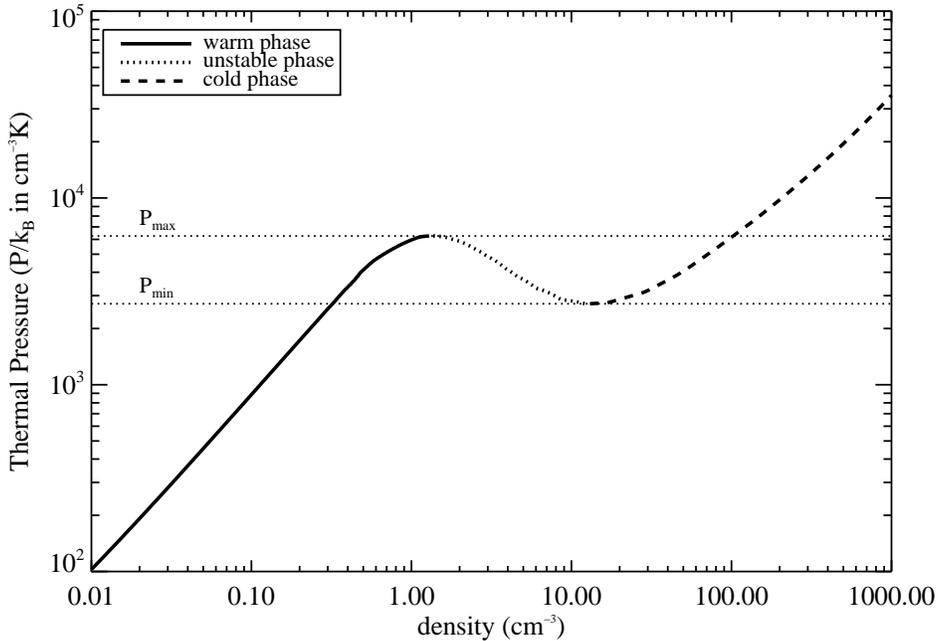}
\caption{Pressure vs.\ density for gas in thermal equilibrium in the ISM.
The warm phase corresponds to gas along the solid portion of the curve while
cold phase gas is for points along the dashed portion.  Gas with
pressure and density putting it on the dotted part of the
curve is unstable and a small displacement from the curve would result in net
cooling or heating, leading to the gas joining either the cold or warm phase.
The dotted horizontal lines indicate the range of pressures for which a stable
two phase medium could exist.\label{fig:thermbal}} 
\end{figure}

Cooling via line radiation and photoelectric heating from gas and dust can
result in a thermally bistable medium within a limited pressure range.
Outside of that range we get only CNM (high $P$) or WNM (low $P$).  This is
illustrated in Figure \ref{fig:thermbal} which makes use of data from
\citet{Wolfire_etal_2003}.  In this diagram gas with $P$ and $n$ placing it at
a point above the curve will tend to cool, while gas below the curve will heat
up.  Thermal conduction tends to make gas move to the left, possibly allowing
gas to make the transition from the cold to the warm phase by overcoming the
effects of radiative cooling.  The hot phase, which would be off the left edge
of the plot, is not a stable phase but has such a long cooling time, that it
effectively behaves like one. The cooling lifetime of hot gas can be
significantly reduced if its density is raised and/or its temperature reduced
through the evaporation or mixing of colder embedded gas clouds.

\begin{figure}[!ht]
\plotone{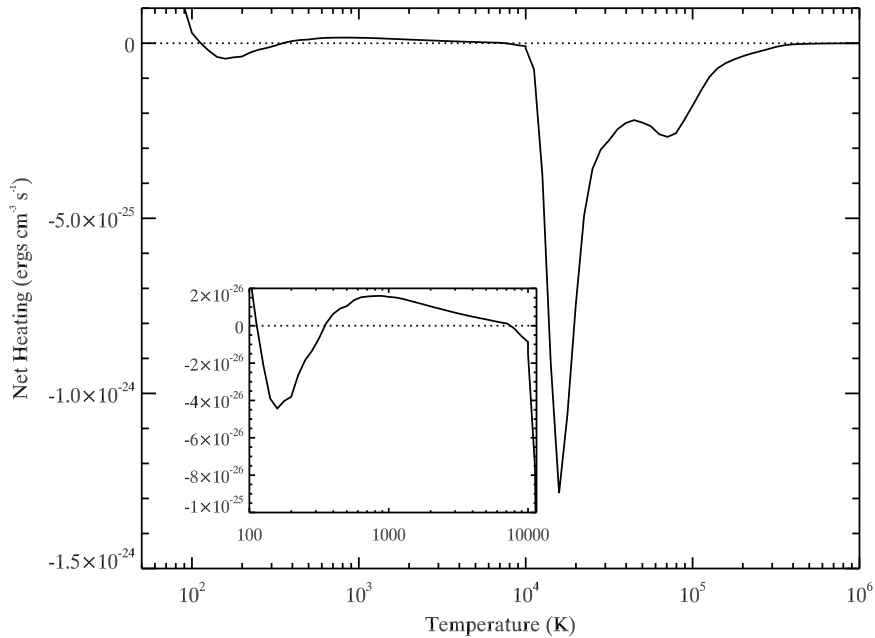}
\caption{Net heating rate of diffuse interstellar gas.  This plot combines the
optically thin cooling rate for hot gas with the net heating rate using data
from \citet{Wolfire_etal_2003} for cold and warm gas and shows how the cooling
rate per unit volume sharply increases above $T \sim 10^4$ K.  Inset is a zoom
on the region between 100 and 10$^4$ K showing more clearly the amplitude of
the variations in the net heating for CNM/WNM temperature
gas.\label{fig:netheat}}
\end{figure}

Gas that is evaporated off from a CNM cloud by thermal conduction will have a
net cooling rate that will increase sharply at first as it is heated.  As its
temperature rises it will cross the instability line in the $P-n$ diagram and
will have net heating.  The net heating goes to zero as the temperature goes
to the WNM temperature.  If the surrounding medium is hot gas, then the
temperature continues to increase and the net cooling increases sharply.  In
this case a large heat flux is needed to balance cooling.  The net heating
curve is illustrated in Figure \ref{fig:netheat}.

\section{Effects of the Ambient Medium on Cloud Evaporation}

CNM clouds surrounded by WNM (rather than hot gas) are protected to some
extent from evaporation. The lower conductivity of neutral gas results in a
lower mass loss rate. Evaporation of cold clouds into warm neutral gas is
simpler than evaporation into hot gas in several ways.  Saturation effects are
not important and for small clouds, $R_\mathrm{cl} \la 0.1$ pc, cooling is
also insignificant.  The evaporation timescale is
\begin{equation}
\tau_\mathrm{evap} = 5.84\times 10^7
\left(\frac{R_\mathrm{cl}}{0.1\,\mathrm{pc}}\right)^2
\left(\frac{T_w}{10^4\,\mathrm{K}}\right)^{-0.8}
\left(\frac{n_\mathrm{cl}}{50\,\mathrm{cm}^{-3}}\right)\;\mathrm{yr}
\end{equation}

CNM clouds immersed in hot gas might be expected to evaporate quickly
\emph{but} as we discussed above, saturation effects are very important for
small clouds. In the high $\sigma_0$ limit (applicable for tiny H\,\textsc{i}
clouds) we get
\begin{equation}
\tau_\mathrm{evap} = 2.2\times 10^6
\left(\frac{R_\mathrm{cl}}{0.1\,\mathrm{pc}}\right)^{7/6}
\left(\frac{n_\mathrm{cl}}{50\,\mathrm{cm}^{-3}}\right)
\left(\frac{P_\mathrm{cl}}{10^4\,\mathrm{cm}^{-3} \mathrm{K}}\right)^{-5/6}\;
\mathrm{yr}
\end{equation}
or expressed in terms of the cloud column
\begin{equation}
\tau_\mathrm{evap} = 8.3\times 10^4 \left(\frac{N_\mathrm{cl}} {10^{18}\,\mathrm{cm}^{-2}}\right)^{7/6}  
\left(\frac{P_\mathrm{cl}} {10^4\,\mathrm{cm}^{-3} \mathrm{K}}\right)^{-1}
\left(\frac{T_\mathrm{cl}} {100\,\mathrm{K}}\right)^{1/6}\; \mathrm{yr}.
\end{equation}
It is interesting to note that the temperature of the hot gas drops out of the
equation for the mass loss rate in this high saturation limit.

\begin{figure}[!ht]
\plotone{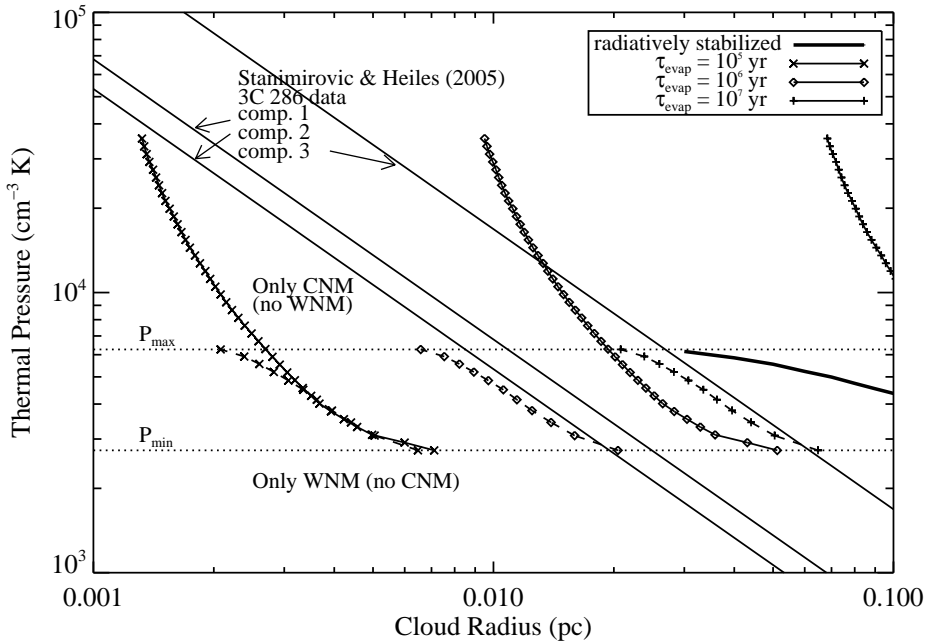}
\caption{Contours of constant evaporation timescales in the $R_\mathrm{cl}-P$
plane. The dashed curves are for CNM clouds embedded in WNM (and extend only
from $P_\mathrm{min}$ to $P_\mathrm{max}$), while the
solid lines are for CNM in hot gas.  The curves connecting the
$\times$'s are for an evaporation lifetime of $10^5$ yr, while the curves with
$\diamond$'s and $+$'s are for lifetimes of $10^6$ and $10^7$ yr respectively.
The thick line shows the $R_\mathrm{cl} - P$ relation for a CNM cloud embedded
in WNM and radiatively stabilized.  The diagonal lines illustrate the
$R_\mathrm{cl} - P$ relation for the tiny H\,\textsc{i} clouds observed by
\citet{Stanimirovich+Heiles_2005}.  The dotted lines show the minimum and
maximum pressures for which a stable two phase medium can exist according to
\citet{Wolfire_etal_2003}.\label{fig:P-Rcl}}
\end{figure}

Depending on the surrounding medium temperature and pressure and the cloud
characteristics, tiny CNM clouds may have short lifetimes against evaporation
or relatively long ones.  In Figure \ref{fig:P-Rcl} we show contours of
constant evaporation lifetime in the $P-R_\mathrm{cl}$ plane.  The plot
includes the case of CNM clouds embedded in hot gas (solid lines) and CNM
embedded in WNM gas (dashed lines).  Clearly the CNM can only evaporate into
WNM if the two phases can co-exist, so the latter lines only run between the
minimum and maximum pressures that allow for a stable two phase medium.  Also
shown are lines derived from the data of \citet{Stanimirovich+Heiles_2005},
who give $N(\mathrm{H\,I})$ and $T_K$ for three components towards 3C 286.
From the Figure it can be seen that if the clouds are embedded in hot gas,
their evaporation lifetime can range from $\sim 5 \times 10^5$ yr to $\sim
2\times 10^6$ yr whereas if they are embedded in WNM the range is $\sim 10^6 -
10^7$ yr.

\section{Discussion}
\begin{figure}[!ht]
\plotone{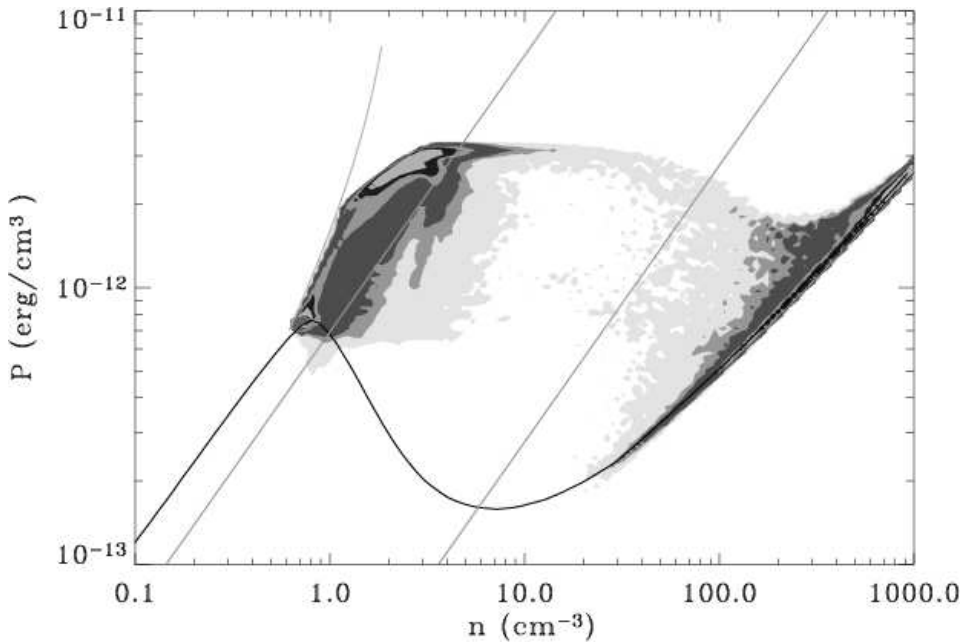}
\caption{Pressure-density diagram color coded to indicate the gas mass
fractions that result after the collision of two WNM clouds
\citep[from][A\&A, 433, 1; reprinted by the kind permission of the authors and
Astronomy \& Astrophysics.  See original article for full color version of
this figure]{Audit+Hennebelle_2005}.  The parallel diagonal lines are
isothermal curves for $T=5000$ K and $T=200$ K. The full black line is the
thermal equilibrium curve. (Note that this curve is somewhat different from
the \citet{Wolfire_etal_2003} equilibrium curve). Much of the mass has been
compressed so as to be out of equilibrium and subject to cooling.  In this way
gas is fed from the WNM to the CNM as a result of the cloud
dynamics.\label{fig:dyn}}
\end{figure}

Recent numerical calculations show that dynamics can produce large amounts of
gas in the ``unstable'' region of the phase diagram.  An example is the work
of \citet{Audit+Hennebelle_2005} in which it is shown that collisions of WNM
clouds create overpressured gas that then cools and becomes CNM gas.  This is
in some sense the inverse process to thermal evaporation.  Figure
\ref{fig:dyn} shows the phase diagram at a time after the interaction of the
two streams is well developed.

Such results lead to the question of the relative importance of thermal
conduction/evaporation versus dynamical processes such as this sort of
triggered condensation.  It is clear that the evaporation timescale is long
enough that evaporation is unlikely to be important in regions experiencing
such strongly dynamical processes.  On the other hand in more quiescent
regions, evaporation could dominate.  The overall importance of conduction
thus depends on an understanding of the larger scale evolution of the ISM and
such things as the frequency of shock crossing for a given shock strength or
the frequency and relative speed for cloud-cloud collisions.  The old
questions of the filling factor of CNM, WNM, WIM (warm ionized medium) and HIM
(hot ionized medium) and their relative arrangement in the ISM clearly also
has bearing on the lifetime for tiny H\,\textsc{i} clouds.  As we have shown,
the ambient medium in which the clouds are embedded could be crucial to
determining their lifetimes.

\section{Conclusions}
Though many may have the impression that small clouds, such as the tiny
H\,\textsc{i} clouds, ought to evaporate very quickly in hot gas, the effects
of heat flux saturation slow the process substantially.  Even without appeal
to tangled magnetic fields the lifetime against evaporation of these clouds
should be on the order of $10^6$ yr.  If the clouds are instead surrounded by
warm neutral gas, \`a la \citet{McKee+Ostriker_1977}, their lifetimes against
evaporation should be increased, though only to $\la 10^7$ yr.  
Thermal evaporation should occur within neutral gas, i.e.\ between CNM and
WNM, and is simpler than between CNM and hot gas since neither the magnetic
field nor radiative cooling nor saturation should play much of a role.  The
one complication in this case is that partial ionization of the warm gas can
substantially reduce its thermal conductivity. More study is needed in this
area, particularly regarding possible observational diagnostics of such
warm-cold evaporation. The final word on the importance of thermal conduction
to the evolution of the tiny H\,\textsc{i} clouds will have to await a more
thorough understanding of the environments in which they exist and the overall
morphology and dynamics of the diffuse ISM.

\acknowledgements
I wish to thank the organizers for inviting me to give this talk and for
organizing such a stimulating conference.  I also thank E.\ Audit and P.\
Hennebelle for granting permission to use their figure and Mark Wolfire for
providing his data on heating-cooling balance in the ISM.  This work was
supported by NASA Astrophysics Theory Program grant no.\ NAG5-13287.

\end{document}